\documentclass{aastex}
\usepackage{emulateapj5}
\usepackage{apjfonts}

\newcommand{\slmc}{s_{\rm LMC}}
\newcommand{\sbkg}{s_{\rm BKG}}
\newcommand{\flmc}{f_{\rm LMC}}
\newcommand{\fbkg}{f_{\rm BKG}}
\newcommand{\fhalo}{f_{\rm MW}}

\shorttitle{The Location of the Source Stars}

\begin{document}

\title{The MACHO Project Hubble Space Telescope Follow-Up:  Preliminary Results on The Location of the Large Magellanic Cloud Microlensing Source Stars}

\author{
      C.~Alcock\altaffilmark{1,2},
    R.A.~Allsman\altaffilmark{3},
      D.R.~Alves\altaffilmark{1,4},
    T.S.~Axelrod\altaffilmark{5},
      A.C.~Becker\altaffilmark{6},
    D.P.~Bennett\altaffilmark{2,7},
    K.H.~Cook\altaffilmark{1,2},
    N.~Dalal\altaffilmark{2,8},
    A.J.~Drake\altaffilmark{1,5},
    K.C.~Freeman\altaffilmark{5},
      M.~Geha\altaffilmark{1,9},
      K.~Griest\altaffilmark{2,8},
    M.J.~Lehner\altaffilmark{10},
    S.L.~Marshall\altaffilmark{1},
    D.~Minniti\altaffilmark{1,11},
    C.A.~Nelson\altaffilmark{1,12},
    B.A.~Peterson\altaffilmark{6},
      P.~Popowski\altaffilmark{1},
    M.R.~Pratt\altaffilmark{6},
    P.J.~Quinn\altaffilmark{13},
    C.W.~Stubbs\altaffilmark{2,14},
      W.~Sutherland\altaffilmark{15},
    A.B.~Tomaney\altaffilmark{14},
      T.~Vandehei\altaffilmark{8} \\
      {\bf (The MACHO Collaboration)}
        }

\altaffiltext{1}{Lawrence Livermore National Laboratory, Livermore, CA 94550\\
    Email: {\tt alcock, kcook, adrake, cnelson, popowski, stuart@igpp.ucllnl.org}}

\altaffiltext{2}{Center for Particle Astrophysics, University of California,
	Berkeley, CA 94720}

\altaffiltext{3}{Supercomputing Facility, Australian National University,
    Canberra, ACT 0200, Australia \\
    Email: {\tt Robyn.Allsman@anu.edu.au}}

\altaffiltext{4}{Space Telescope Science Institute, 3700 San Martin Dr.,
    Baltimore, MD 21218\\
    Email: {\tt alves@stsci.edu}}

\altaffiltext{5}{Research School of Astronomy and Astrophysics,
        Canberra, Weston Creek, ACT 2611, Australia\\
 Email: {\tt tsa, kcf, peterson@mso.anu.edu.au}}

\altaffiltext{6}{Bell Laboratories, Lucent Technologies, 600 Mountain Avenue, 
	Murray Hill, NJ 07974\\
  Email: {\tt acbecker@physics.bell-labs.com}}

\altaffiltext{7}{Department of Physics, University of Notre Dame, IN 46556\\
    Email: {\tt bennett@bustard.phys.nd.edu}}

\altaffiltext{8}{Department of Physics, University of California,
    San Diego, CA 92039\\
    Email: {\tt endall@physics.ucsd.edu, kgriest@ucsd.edu, vandehei@astrophys.ucsd.edu }}

\altaffiltext{9}{Department of Astronomy and Astrophysics,
     University of California, Santa Cruz 95064 \\
     Email: {\tt mgeha@ucolick.org}}

\altaffiltext{10}{Department of Physics, University of Sheffield, Sheffield S3 7RH, UK\\
    Email: {\tt m.lehner@sheffield.ac.uk}}

\altaffiltext{11}{Depto. de Astronomia, P. Universidad Catolica, Casilla 104,
        Santiago 22, Chile\\
Email: {\tt dante@astro.puc.cl}}

\altaffiltext{12}{Department of Physics, University of California, Berkeley,
        CA 94720}

\altaffiltext{13}{European Southern Observatory, Karl Schwarzchild Str.\ 2,
        D-8574 8 G\"{a}rching bel M\"{u}nchen, Germany\\	
	Email: {\tt pjq@eso.org}}

\altaffiltext{14}{Departments of Astronomy and Physics,
    University of Washington, Seattle, WA 98195\\
    Email: {\tt stubbs@astro.washington.edu}}

\altaffiltext{15}{Department of Physics, University of Oxford,
    Oxford OX1 3RH, U.K.\\
    Email: {\tt w.sutherland@physics.ox.ac.uk}}

\begin{abstract}
We attempt to determine whether the MACHO microlensing source stars are drawn
from the average population of the LMC or from a population behind the LMC by
examining the HST color-magnitude diagram (CMD) of microlensing source stars.
We present WFPC2 HST photometry of eight MACHO microlensing source stars and
the surrounding fields in the LMC.  The microlensing source stars are
identified by deriving accurate centroids in the ground-based MACHO images
using difference image analysis (DIA) and then transforming the DIA
coordinates to the HST frame.  We consider in detail a model for the
background population of source stars based on that presented by Zhao, Graff
\& Guhathakurta.  In this model, the source stars have an additional reddening
$<E(B-V)>=0.13$ mag and a slightly larger distance modulus $<\Delta \mu> \sim
0.3$ mag than the average LMC population.  We also investigate a series of
source star models, varying the relative fraction of source stars drawn from
the average and background populations and the displacement of the background
population from the LMC.  Due to the small number of analyzed events the
distribution of probabilities of different models is rather flat.  A shallow
maximum occurs at a fraction $\slmc \sim 0.8$ of the source stars in the LMC.
This is consistent with the interpretation that a significant fraction of
observed microlensing events are due to lenses in the Milky Way halo, but
does not definitively exclude other models.
\end{abstract}


\section{Introduction}

The crucial observable in microlensing, the event duration, admits degeneracy
in the three fundamental microlensing parameters: the mass, distance and
velocity of the lens.  This makes it difficult to distinguish between the two
principal geometric arrangements which may explain Large Magellanic Cloud
(LMC) microlensing: a) MW-lensing, in which the lensed object is part of the
Milky Way (MW) and b) self-lensing, in which the lensed object is part of the
LMC.  In self-lensing, the lens may belong to the disk+bar, halo or ``shroud''
of the LMC, while the source star may come from any of these components or
some sort of background population to the LMC.

Most efforts to distinguish between MW-lensing and self-lensing event
distributions focus on modeling the LMC self-lensing contribution to the
optical depth and comparing this to the observed optical depth.  Early works
by \citet{sah94} and \citet{wu94} considered the traditional LMC self-lensing
geometry in which both the source and lens are in the disk+bar of the LMC.
Both works suggested that disk+bar self-lensing could account for a
substantial fraction of the observed optical depth.  This claim has since been
disputed by several other groups \citep{gou95,alc97,alc00a} who show that when
considering only disk stars the rate of LMC self-lensing is far too low to
account for the observed rate.  \citet{gyu00} show that allowing for
contributions to the lens and source populations from the LMC bar does not
substantially increase the LMC self-lensing optical depth. \citet{alv00a} find
a low LMC self-lensing optical depth for a flared LMC disk. 
\citet{gyu00} also show that much of the disagreement in models of
the disk+bar self-lensing optical depth results from disagreement about the
fundamental parameters of the LMC, such as the total disk mass and inclination
angle.  Within their region of allowed parameters \citet{gyu00} also make a
strong case that disk+bar self-lensing makes a small contribution to the observed
optical depth, at most $\sim$20\%.

However, self-lensing becomes a much more plausible hypothesis if one
allows for lenses in an LMC stellar halo population.  The principal problem
surrounding an LMC stellar halo contribution is that no tracers of old populations
in the LMC have ever revealed a population with high enough velocity
dispersions to suggest a \textit{virialized} spheroidal component \citep{ols96}.  

Recently however, a new possibility has arisen from the results of
\citet{wei00} who claims that LMC microlensing may be caused by a
\textit{non-virialized} stellar halo or ``shroud''.  This term was introduced
by \citet{eva00} and is meant to imply an LMC population which is like a halo
in that it is spacially not part of the LMC disk, but unlike a halo in that it
is non-virialized and thus may have a relatively low velocity dispersion.
Such a population is suggested by the simulations of \citet{wei00} who finds
that the LMC's dynamical interaction with the MW may torque the LMC disk in
such a way that the LMC disk is thickened and a spheroid component is
populated without isotropizing the stellar orbits and thereby leaving disklike
kinematics intact.  Statistically marginal evidence for such a kinematically
distinct population is found observationally in a study of carbon star
velocities by \citet{gra00}.  

However, observational work on RR Lyrae by \citet{kin91}, which does not rely
on the specific kinematics of the spheroidi, limits the total mass of any type of
halo (virialized or non-virialized) to perhaps 5\% of the mass of the LMC, too
small to contribute more than $\sim$5\% of the observed optical depth.  A more
recent revisiting of this argument by \citet{alv00b} finds room to increase
this optical depth contribution to at most 20\%, still only a small fraction
of the total. 

In order for an LMC shroud to account for the total optical depth it must
have a mass comparable to that of the LMC disk+bar \citep{gyu00}.  
Even if we accept the existence of such a massive shroud, the
microlensing implications are somewhat in dispute.  \citet{wei00} finds an LMC
self-lensing optical depth comparable to the observed optical depth.  However,
this estimate is reduced by a factor of three by \citet{gyu00} who repeat the
\citet{wei00} microlensing analysis using lower values for the disk total mass
and inclination angle and a proper weighting over all observed MACHO fields.

Yet another self-lensing geometry was introduced by \citet{zha99} who suggests
that the observed events are due to ``background'' self-lensing in which the
source stars are located in some background population, displaced at some
distance behind the LMC. A veritable plethora of lenses for this population is
then supplied by the disk+bar of the LMC. A background population has the
advantage of being nearly impossible to confirm or reject observationally, as
there are nearly no limits on its size or content (provided of course, it is 
small enough to ``hide'' behind the LMC).

A final possibility is ``foreground'' self-lensing.  This is not self-lensing
in the classical sense as in this case the lenses are not drawn from the LMC
itself, but rather from some kinematically distinct foreground population,
such as an intervening dwarf galaxy.  \citet{zar97} claim a detection of a
population of stars from such an entity.  However, \citet{bea98} claim that
this ``population'' is a morphological feature of the LMC red clump, while
others show that such a population consistent with other observational
constraints could not produce a substantial microlensing signal
\citep{gou98,ben98}.

In this work we attempt to determine whether the MACHO source stars belong to
the average population of the LMC or to a background population displaced at
some distance behind the LMC disk. The determination of source star location
is based on the suggestion of \citet{zha99}, \citet{zha00a}, and
\citet{zha00b} who point out that source stars from a background population
should be preferentially fainter and redder than the average population of the
LMC due to the extinction of the LMC disk and their displacement along the
line of sight.  \citet{zha00b} present a model for this background population
with an additional mean reddening relative to the average population of the
LMC $<E(B-V)> = 0.13$ mag and a displacement from the LMC of $\sim 7.5$ kpc
resulting in an increase of distance modulus of $<\Delta \mu> \sim 0.3$ mag. 

The location of the source stars has implications for the location of the
lenses and thus for the nature of LMC microlensing.  If all the source stars
are in the background population, then the great majority of the lenses are
found in the LMC disk+bar and LMC microlensing is dominated by background
self-lensing.  Conversely, if all the source stars are in the LMC, then 
microlensing may be due to MW-lensing, disk+bar self-lensing or foreground
self-lensing.  However, since the contribution from disk+bar self-lensing has
been shown to be small and the evidence for a foreground intervening
population is unconvincing, a result which places all source stars in the LMC
would suggest that LMC microlensing is dominated by MW-lensing.  
If, however, we find a more equal division of source stars between the LMC and
the background population then this implies either some mixture of MW-lensing
and disk+bar self-lensing, or a more symmetric self-lensing geometry such as the
LMC shroud discussed above.

We first investigate two models: the first putting all source stars in the LMC
(Model 1), and the second putting all source stars in a background population
(Model 2). We compare a Hubble Space Telescope (HST) CMD of MACHO microlensing
source stars to efficiency weighted CMDs of the average population of the LMC
and the \citet{zha00b} background population. In \S 2 we construct a CMD of
the average LMC population by combining the CMDs of eight HST Wide Field
Planetary Camera 2 (WFPC2) fields centered on past MACHO microlensing events
in the outer LMC bar.  In \S 3 we describe the identification of the
microlensing source stars in these fields by difference image analysis (DIA).
In \S 4 we construct the background population CMD by shifting the HST CMD by
the appropriate amount of extinction and distance modulus.  We then describe
the convolution of the average and background HST CMDs with the MACHO
efficiency for detecting a microlensing event in a source star of given
magnitude. In \S 5 we determine the likelihoods that the microlensing source
stars were drawn from the average population of the LMC (Model 1) and from a
population of background source stars (Model 2) by using Kolmogorov-Smirnov
(KS) tests to compare our observed and model distributions.  Finally, in \S 6
we generalize our analysis and consider intermediate models with varying
distance moduli and fractions of source stars in the background and LMC. We
conclude and discuss the implications for the location of the lenses in \S 7.

\section{HST Observations}

Observations were made with the WFPC2 on HST between May 1997 and October 1999
through the F555W ($V$) and F814W ($I$) filters.  The Planetary Camera (PC)
was centered on the location of past MACHO microlensing events.  The
microlensing events, positions, and exposure times are listed in Table 1.

Multiple exposures of a field were combined using a sigma-clipping algorithm
to remove deviant pixels, usually cosmic rays.  The PC has a pixel size of
$0.046''$ which easily resolves the great majority of stars in our frames.
Most stars are also resolved in the Wide Field (WF) fields which have a pixel
size of $0.1''$.  Instrumental magnitudes were calculated from aperture
photometry using DAOPHOT II \citep{ste87,ste91} with a radius of $0.25''$ and
centroids derived from point-spread function (PSF) fitting photometry.
Aperture corrections to $0.5''$ were performed individually for each frame. We
correct for the WFPC2 charge transfer effect using the equations from
Instrument Science Report WFPC2 97-08.  We also make the minimal corrections
for contaminants which adhere to the cold CCD window according to the WFPC2
Instrument Handbook.  We transform our instrumental magnitudes to Landolt $V$
and $I$ using the calibrations from \citet{hol95}.

\begin{figure}
\epsscale{0.5}
\plotone{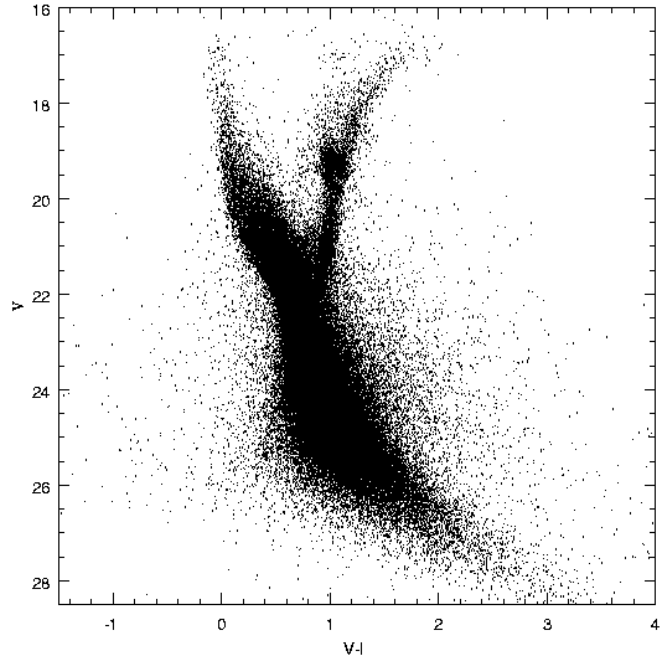}
\caption{The composite HST CMD created by combining the
photometry from 8 WFPC2 fields centered on observed MACHO microlensing events.}
\label{hst}[f1.ps]
\end{figure}

We create a composite LMC CMD by combining the PC and WF photometry for all of
our fields except the field of LMC-1.  In the case of LMC-1 the $V$ and $I$
observations were taken at different roll angles and there is little area of
overlap except in the PC frame.  We therefore include the PC field from LMC-1
but not the WF fields.  The composite HST CMD is shown in Figure~\ref{hst}.

\section{Source Star Identification through Difference Image Analysis} 

A ground-based MACHO image has a pixel size of $0.6''$ and a seeing of at
least $1.5''$.  Thus, in a typically crowded region of the outer LMC bar, a
MACHO seeing disk will contain $\sim$11 stars of $V\lesssim24$.  This means that
faint ``stars'' in ground-based MACHO photometry are usually not single stars at
all, but rather blended composite objects made up of several fainter stars.
Henceforth, we distinguish between these two words carefully, using
\textit{object} to denote a collection of stars blended into one seeing disk,
and \textit{star} to denote a single star, resolved in an HST image or through
DIA.  The characteristics of the MACHO object that was lensed tell us little
about the actual lensed star.  However, with the microlensing object centroid
from the MACHO images we can hope to identify the microlensing source star in
the corresponding HST frame.

\begin{figure}
\epsscale{0.5}
\plotone{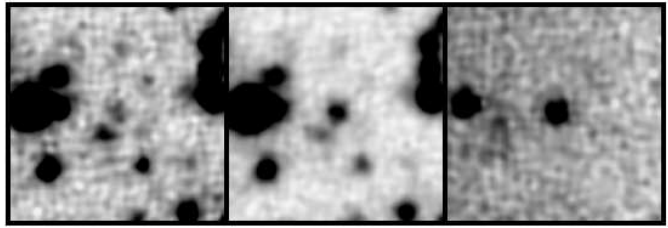}
\caption{The left panel shows a 0.6' X 0.6' section of the
  baseline image of MACHO event LMC-4. The middle panel
  shows the same region taken at the peak of the microlensing event.  The 
  right panel shows the differenced image.  The flux at the left hand
  side of the differenced image is due to an asymptotic giant branch variable
  star at that location. }
\label{macho}[f2.ps]
\end{figure}

\begin{figure}
\epsscale{0.5}
\plotone{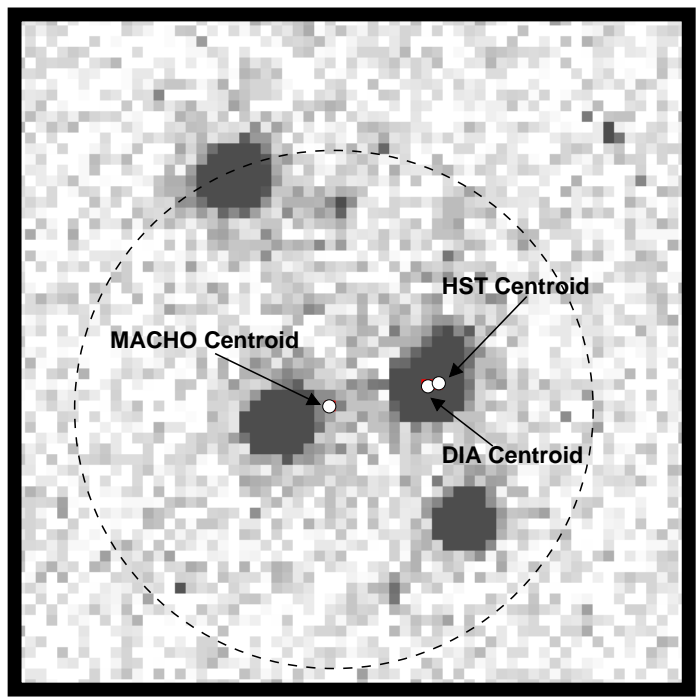}
\caption{A 3'' X 3 '' HST image of LMC-4.  The circle contains the
several HST stars which are all contained within the MACHO seeing disk of the
lensed object.  The arrows indicate the MACHO baseline centroid, the DIA
centroid and the HST centroid. }
\label{lmc4}[f3.ps]
\end{figure}

A direct coordinate transformation from the MACHO frame to the HST frame often
places the baseline MACHO object centroid in the middle of a cluster of faint
HST stars with no single star clearly identified.  To resolve this ambiguity we
have used DIA.  This technique is described in detail in \citet{tom96}, but we
review the main points here.  DIA is an image subtraction technique designed
to provide accurate photometry and centroids of variable stars in crowded
fields.  The basic idea is to subtract from each program image a high
signal-to-noise reference image, leaving a differenced image containing
only the variable components. Applied to microlensing, we subtract baseline
images from images taken at the peak of the microlensing light curve, leaving
a differenced image containing only the flux from the microlensing source
star and not the rest of the object. We also find a centroid shift between the
baseline image and the differenced image towards the single star that was
microlensed.  If the centroid from the differenced image is transformed to the
HST frame we find that it usually clearly identifies the HST microlensed
source star.  This process is illustrated in Figures~\ref{macho} and
\ref{lmc4}.

This technique allows us to unambiguously identify 7 of 8 microlensed source
stars.  In the case of LMC-9 the DIA centroid lands perfectly between two
stars; fortunately these two stars are virtually identical sub-giants and the
choice between the two has no effect on our results.  In Table 2 we present
the $V$ magnitudes and $(V-I)$ colors of our source stars from the HST data.
The errors presented here are the formal photon counting errors returned by
DAOPHOT II.  We estimate that all WFPC2 magnitudes have an additional
$0.02-0.03$ mag uncertainty due to aperture corrections.  In the case of LMC-9
we tabulate both possibilities and use LMC-9a in the remainder of this work.

Our identification of LMC-5 revealed it to be the rather rare case of a
somewhat blended HST star.  Although there are two stars evident, at an
aperture of $0.25''$ the flux of one star was contaminated by that of its
neighbor. Therefore, in this case, we perform PSF fitting photometry using
PSFs kindly provided by Peter Stetson.  The errors presented in Table 2 for
LMC-5 are those returned by the profile fitting routine ALLSTAR. The DIA
centroid falls 2 pixels closer to the centroid of star one than star two,
clearly preferring star one as the source star. Furthermore, as predicted by
\citet{alc97} and \citet{gou97}, star two is a rather red object which is very
faint in the V band.  Fits to the MACHO lightcurve presented in \citet{alc00a}
suggest lensed flux fractions in the V and R bands of 1.00 and 0.46
respectively, confirming the DIA choice of the much bluer star as the lensed
source star. 

\begin{figure}
\epsscale{0.5}
\plotone{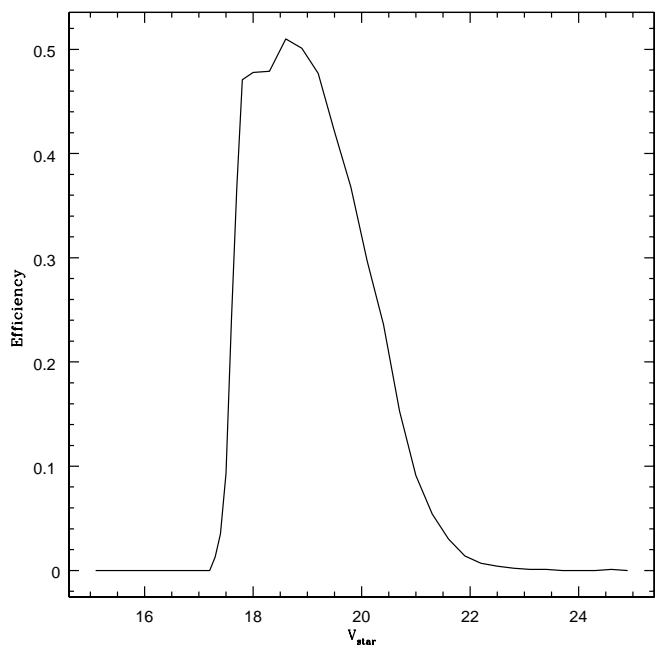}
\caption{The MACHO detection efficiency as a function of stellar V-magnitude.
That is, if a microlensing event occurs in a \textit{star} of given magnitude
$V_{star}$, this is the given efficiency for detecting that event.}
\label{eff}[f4.ps]
\end{figure}

Since this work specifically addresses the background lensing geometry, we
also discuss here the (remote) possibility that our HST images of the source
stars are actually completely blended objects consisting of a faint background
source star and a brighter LMC lens.  Such a configuration would seriously
skew our CMD distribution of source stars as we would instead be presenting
photometry of the lenses.  We begin by noting that the MACHO efficiency for
detection of a microlensing event in a \textit{star} has fallen to zero at $V
\sim 22.5$ (see Figure~\ref{eff} and explanation below).  This means that a
``faint'' background source star must have $V_S < 22.5$ in order to produce a
detectable event.  The the lens in this scenario is assumed to have $V_L <
V_S$. We estimate that we would not recognize a blended object of two stars
with $V < 22.5$ as such if the centroids coincided to within 1.5 pixels. In
our most crowded PC field we find $\sim$1000 stars with $V < 22.5$ spread over
an area of 720 X 720 pixels. In simulations, we draw 1000 stars with $V <
22.5$ weighted according to our luminosity function and spread randomly over
720X720 pixels.  For each star we then check to see if it is found within 1.5
pixel of a brighter star. We find that on average, there will be 5 stars of $V
< 22.5$ which are blended with a brighter star.  Therefore, for our most
crowded field, the chance that our source star is an unrecognized blend of a
faint source star and a brighter lens is about 5 in a thousand. In a more
typically crowded field, this falls to around 1 chance in 1000.  Therefore, it
is extremely unlikely that any of our 8 HST source stars are blended objects
composed of a faint source star and a bright lens.

\section{Creation of the model source star populations}

If the microlensing events are due to MW lenses, then one would expect the
distribution of observed microlensing source stars to be randomly drawn from
the average population of the LMC corrected only for the MACHO detection
efficiency for stars of a given magnitude (Model 1).  We assume that the
population of the LMC is well represented by our composite HST CMD to
$V\lesssim24$.  If the microlensing events are background self-lensing events
we expect the source stars to be drawn from a background population which
suffers from the internal extinction of the LMC (Model 2).  To represent such
a background population we shift the composite HST CMD according to the
amounts suggested by \citet{zha00b}, $<E(B-V)>=0.13$ and $\Delta \mu = 0.3$.
Since, \citet{hol95} calibrate instrumental WFPC2 magnitudes to the Landolt
system, we use the appropriate Landolt system extinction coefficients of Table
6 in \citet{sch98} to translate these estimates to our filters.  The total
shifts, taking into account both reddening and distance modulus, are

\begin{displaymath}
\Delta V = A_{V} + \Delta{\mu} = 0.73, ~~ \Delta(V-I) = E(V-I) =
0.18~~~~~~~(Model~2)
\end{displaymath}

\begin{figure}
\epsscale{0.5}
\plotone{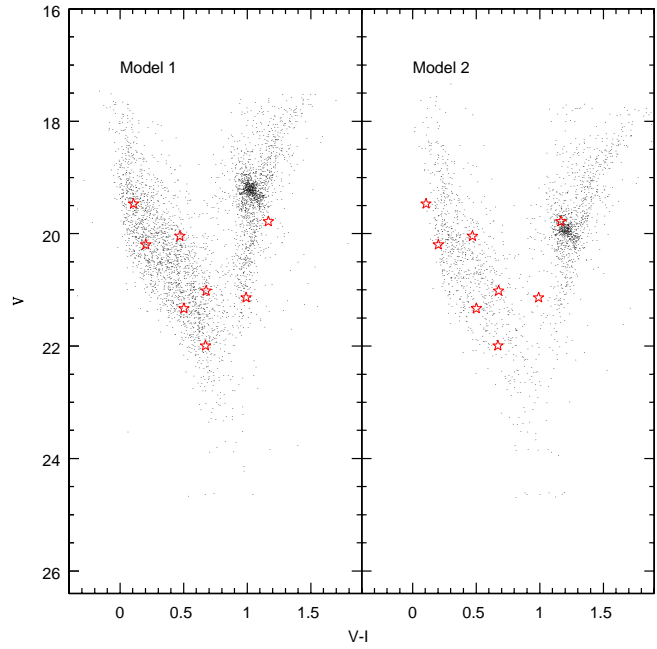}
\caption{The model source star populations.  Model 1 represents a source star
population in the disk of the LMC, Model 2 represents a source star population
behind the LMC.  The MACHO microlensing events of Table 2 are overplotted in
red.  We perform 2-D KS tests to determine the probability that the
microlensing events are drawn from each model source star populations.  We
find probabilities $P_{1}=0.319\pm0.027$ and $P_{2}=0.103\pm0.023$ that the
microlensing events are consistent with the source star populations of Models
1 and 2, respectively.} \label{cmd}[f5.ps]
\end{figure}

Thus far we have constructed two CMDs representing the distribution of all
possible source stars down to $V\sim24$.  However, not all possible
microlensing events are detected in the MACHO images.  To create a CMD
representing a population of source stars which produce detectable
microlensing events we must convolve the HST CMD with the MACHO detection
efficiency. The MACHO efficiency pipeline is extensively described in
\citet{alc00a} and \citet{alc00b} and the detection efficiency as a function
of stellar magnitude, $V_{\rm{star}}$, and Einstein ring crossing time has
been calculated.  We average this function over the event durations of the
candidate microlensing events derived using detection criterion A from \citet{alc00a}
and present the MACHO detection efficiency as a function of $V_{\rm{star}}$ in
Figure~\ref{eff}.  We convolve this function with our HST CMDs to produce the
final Model 1 and 2 distributions of source stars.  In Figure~\ref{cmd}, we
show our model source star populations with the observed microlensing source
stars of Table 2 overplotted as large red stars.

This procedure admits several assumptions.  First, we assume that our eight
HST fields collectively well represent the stellar population of the LMC disk.
This assumption has two parts, the first being that an observation at a random
line of sight in the LMC bar is dominated by stars in the LMC disk and the
second that the stellar population across the LMC is fairly constant.  The
first part holds so long as the surface density of the background population
is much smaller than that of the LMC itself.  If this were not the case, this
population would have been directly detected. The second part has been
confirmed by many LMC population studies including
\citet{alc00b}, \citet{ols99}, and \citet{geh98}, as well as our own comparison
of individual CMDs and luminosity functions.  Second, we assume that
the underlying stellar content of the background population is identical to
that of the LMC. 

\section{The Location of the Source Stars}

We now attempt to determine whether the CMD of microlensed source stars is
consistent with the average population of the LMC (Model 1) or whether it is
more consistent with a background population (Model 2) by performing a
two-dimensional Kolmogorov-Smirnov (KS) test.

In the familiar one dimensional case, a KS test of two samples with number of
points $N_{1}$ and $N_{2}$ returns a distance statistic $D$, defined to be the
maximum distance between the cumulative probability functions at any ordinate.
Associated with $D$ is a corresponding probability $P(D)$ that \textit{if} two
random samples of size $N_{1}$ and $N_{2}$ are \textit{drawn from the same
distribution} a worse value of $D$ will result.  This is equivalent to saying
that we can exclude the hypothesis that the two samples are drawn from the
same distribution at a confidence level of $1.0-P(D)$. If $N_{2} \gg N_{1}$ then
this is also equivalent to excluding at a $1.0-P(D)$ confidence level the
hypothesis that sample 1 is drawn from sample 2.

The concept of a cumulative distribution is not defined in more than one
dimension.  However, it has been shown that a good substitute in two
dimensions is the integrated probability in each of four right-angled
quadrants surrounding a given point \citep{fas87,pea83}.  Leaving
aside the exact algolrithmic definition \citep{pre92} a two-dimensional
KS test yields a distance statistic $D$ and a corresponding $P(D)$ with the
same interpretation as in the one dimensional case.

We use the two-dimensional KS test to test hypothesis that the CMD of observed
MACHO microlensing events is drawn from the same population as each of the
model source star distributions.  We find distance statistics $D_1 = 0.394\pm
0.005$ and $D_2 = 0.473\pm0.009$, for Models 1 and 2 respectively.  Each of
these distance statistics has a corresponding probability, $P(D)$, that if we
draw an 8 star samples from the model population a larger value than $D$ will
result.  As explained above, this is equivalent to excluding this model
population as the actual parent of our observed microlensing source stars at a
confidence level of $1.0-P(D)$.  These probabilities are $P_1 =0.319\pm0.027$
and $P_2=0.103\pm0.023$.  The error quoted for each of these quantities is the
scatter about the mean value in 50 simulations for each model.  Because the
creation of the efficiency convolved CMD is a weighted random draw from the
HST CMD, the model population created in each simulation differs slightly.
This in turn leads to small differences in the KS statistics.

These results tell us that these 8 MACHO events are insufficient to reliably
distinguish the MW and self lensing hypothesis. The best we can do is to
exclude Model 2 at the statistically marginal 90\% confidence level.  

\section{Intermediate Models}

Thus far we have considered the possibilities that the observed MACHO source
stars are either all LMC stars or all background stars at a mean distance of
$\sim7.5$ kpc behind the LMC.  However, as discussed in \citet{zha99},
\citet{zha00a} and \citet{zha00b} there is substantial middle ground.  In a
complete analysis, we may treat both the fraction of source stars drawn from
the background population and the distance to the background population as
adjustable parameters. While the size, location and content of the LMC has
been well constrained by observations, the existence, size and location of a
background population is constrained only by the fact that it must be small
enough to have evaded direct detection. The distance to the background
population from \citet{zha00b} is very loosely derived by the requirement that
the background population be at least transiently gravitationally bound to the
LMC.  However, the reddening of a background population is a much more
physically constrained number since a population behind the LMC should certainly
suffer from the mean internal extinction of the LMC, a number which has been
well determined in a number of studies including \citet{oes96} and
\citet{har97}.  Therefore, all our background population models have the same
reddening $<E(B-V)>=0.13$, as inferred from the mean extinction of the LMC
from \citet{har97} corrected for Galactic foreground extinction.

We define $\slmc$ to be the fraction of the source stars drawn from the LMC
disk+bar population, leaving a fraction $1.0-\slmc$ source stars drawn from the
background population, and $\Delta \mu$ to be the excess distance modulus of
the background population.  We consider values $\Delta \mu = 0.0, 0.30, 0.45$
and for each value of the distance modulus we consider the full range of
$\slmc$ from 0.0 to 1.0.  For example, a model with $\Delta \mu = 0.45$ and
$\slmc=0.5$ contains a mixture of source stars in which half the source stars
are drawn from the population of the LMC disk, and half the source stars are
drawn from a background population displaced from the LMC by $\sim11$kpc and
reddened by $<E(B-V)>=0.13$.  All models with $\slmc=0.0$ contain only source
stars drawn from a specified background population, and all models with
$\slmc=1.0$ are identical, containing only stars drawn from the LMC disk.

\begin{figure}
\epsscale{0.5}
\plotone{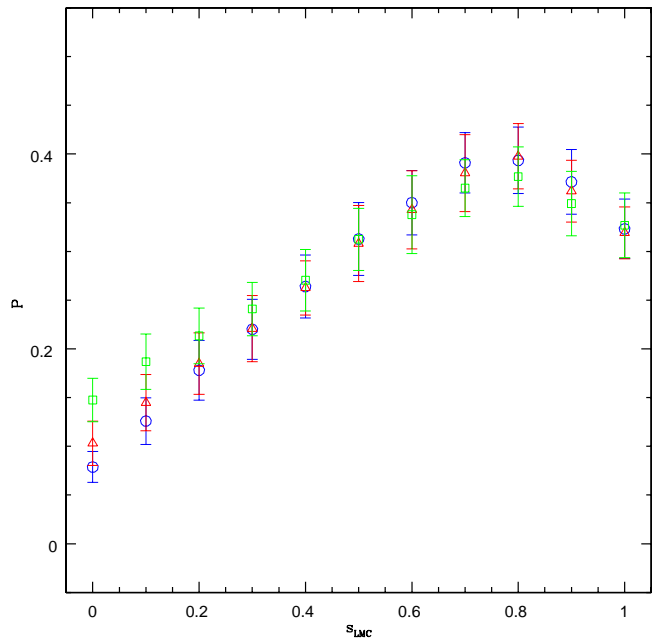}
\caption{The KS test probabilities, P, for various values of the fraction of
source stars in the LMC, $\slmc$,  and the displacement of the background
population, $\Delta \mu$. We show the results for $\Delta \mu = 0.45,0.30,0.0$
in blue circles, red triangles and green squares, respectively.  The error
bars indicate the scatter around the mean value of 50 simulations done
for each model.} 
\label{fmacho}[f6.ps] 
\end{figure}

We present the results in Figure~\ref{fmacho}, showing the KS test
probabilities as a function of $\slmc$ for each of our values of $\Delta
\mu$. Again, the error bars reflect the scatter about the mean for
fifty simultations of each model. We note that since all models with
$\slmc=1.0$ contain only source stars drawn from the LMC disk, all curves
for different $\Delta \mu$ must converge at $\slmc=1.0$.  Furthermore, we
learn from Figure~\ref{fmacho} that the value of $\Delta \mu$ makes little
difference even at $\slmc=0.0$. So long as the background population is
behind the LMC and therefore reddened by $<E(B-V)>=0.13$ its exact
displacement is of little import.  In all cases a smaller value of $\Delta
\mu$ flattens the curves somewhat.  This is expected as a smaller value of
$\Delta \mu$ implies overall less difference between the two extreme models at
$\slmc=0.0$ and $\slmc=1.0$. The two dimensional KS test of the CMD
displays a shallow maximum at $\slmc \sim0.8$.

\section{Discussion}

We have compared two models for LMC microlensing: source stars drawn from the
average population of the LMC and source stars drawn from a population behind
the LMC.  By comparing the CMD of observed microlensing source stars to a CMD
representing all detectable microlensing events for each of these models we
find by two-dimensional KS tests that the data suggest that it is more likely
that all the source stars are in the LMC than that all the source stars are in
the background.  However, we can only exlude the possibility that all the
source stars are in the background at a statistically marginal 90\% confidence
level.

We also consider a number of intermediate models in which we vary the distance
modulus of the background population as well as the fraction of stars drawn
from average and background populations.  In these models we find that for all
displacements the most highest probability occurs for a fraction $\slmc \sim
0.8$ LMC source stars and $\sbkg= 1.0 - \slmc = 0.2$ background source stars.
We also find that the value of the distance modulus has very little effect on
our results at all values of $\slmc$.

Our results are completely consistent with the interpretation that MACHO
microlensing events are dominated by MW-lensing.  In the MW-lensing geometry,
the source stars reside in the LMC and the lenses are in the halo or disk of
the MW.  The MW-lensing interpretation for microlensing requires that $\slmc
\sim 1.0$, consistent with our result.  We note that both disk+bar
self-lensing and foreground self-lensing also place all their source stars in
the LMC.  However, these contributions to the number of events have been
shown to be small and we neglect their contribution here.  We also note that
the contribution from the known stellar populations of the MW is expected to
be small ($\sim$1 events of the 13-event cut A sample of \citet{alc00a} but
not entirely negligible.  A lens population in the MW is likely to be
dominated by MACHOs in the MW halo and not faint stars in the MW disk
or spheroid.

\citet{zha99}, \citet{zha00a} and \citet{zha00b} propose a model for LMC
microlensing in which all the source stars are drawn from some background
population to the LMC.  Such a model suggests that LMC microlensing is
dominated by background self-lensing and that $\slmc \sim 0.0$.  We find this
arrangement to be the model excluded at the highest confidence; however, we
cannot rule it out with any great statistical weight.

Recently, \citet{wei00} and \citet{eva00} have suggested a model of LMC
microlensing in which all microlensing events are due to a non-virialized
shroud of stars which surrounds the LMC.  In shroud self-lensing there are
four event geometries: a) background shroud source and disk lens, b) disk
source and foreground shroud lens, c) disk source and disk lens and d)
background shroud source and foreground shroud lens.  We might naively expect
the former two types dominate the number of expected events and so if we were
to ignore the contribution from the latter two types we would conclude that
shroud lensing would imply $\slmc \sim 0.5$.  However, in order to produce the
entire observed optical depth, the shroud must be so massive that it is no
longer self-consistent to ignore the latter terms.  Calculations performed in
the formalism of \citet{gyu00} suggest instead that events with a background
shroud source and a foreground shroud lens become an important contributor and
reduce the expected fraction of source stars in the LMC to $\slmc \sim 0.3 -
0.4$.  Our results are not inconsistent with such a model, however we note
that there is a profound lack of observational evidence for such a stellar
shroud.

Furthermore, if we assume that a substantial LMC stellar halo or shroud is not
a realistic possibility, then we may directly relate the fraction of source
stars in the LMC, $\slmc$, to the fraction of microlensing events which are
MW-lensing, $\fhalo$ in the following way.  All events in which the source
stars are located in the background are, by definition, background lensing
events.  Therefore, $\sbkg = \fbkg$, where $\fbkg$ indicates the fraction of
observed events which are background lensing.  The total fraction of lensing
events due to halo-lensing $\fhalo$, LMC self-lensing, $\flmc$, and background
self-lensing, $\fbkg$, must be equal to unity.
\begin{equation}
\fhalo + \flmc + \fbkg = 1.0
\end{equation}
This may be rearranged to read
\begin{equation}
\fhalo = 1.0 - \fbkg - \flmc = \slmc - \flmc
\end{equation}
Equation (2) is strictly true even if we allow for LMC shroud lensing.  However,
if we ignore the possibility of LMC self-lensing by a non-virialized stellar
shroud, then all models with reasonable parameters for the LMC find that
$\flmc \lesssim 0.2$.  Therefore, if $\slmc \sim 1$, we may make the
approximation
\begin{equation}
\fhalo \sim \slmc
\end{equation}
Therefore we estimate, with low statistical significance due to the small
sample size of our sample, that a fraction $\fhalo \sim 0.8$ of observed microlensing
events are halo-lensing.  We emphasize once again that this conclusion ignores
the possibility of a non-virialized LMC shroud.

At present, the strength of this analysis is severely limited by the number of
microlensing events for which we have corresponding HST data.  A fortuitous
distribution in the CMD of all 13 criterion A events presented in
\citet{alc00a} may allow us to definitively exclude a model in which all the
source stars are drawn from the background population.  It is difficult to
estimate how many events are needed to definitely determine $\slmc$ as the
number depends on the true value of $\slmc$ as well the degree of accuracy one
wishes to achieve.  However, when we perform Monte Carlo simulations where we
draw a sample of N events from the distribution with $\slmc = 0.0$ and compare
it with the 2-D KS test to the distribution with $\slmc = 1.0$ we find that
for $N \sim 20 - 25$, $P < 0.01$ for at least 99\% of our simulations.  This
implies that if the true value of $\slmc$ is near either extreme ($\slmc \sim
0$ or $\slmc \sim 1$) then a CMD of 20-25 events virtually guarantees that we
will be able to exclude a model at the other extreme at the 99\% confidence
level.  Even more events are necessary to exclude intermediate models with
various fractions of LMC and background stars.  Ongoing microlensing search
projects (EROS, OGLE II) may supply a sufficient sample of events in the next
few years.  The technique outlined in this paper should prove a powerful
method for locating the lenses with these future datasets.

\acknowledgments
Support for this publication was provided by NASA through proposal numbers
GO-5901 and GO-7306 and from the Space Telescope Science Institute, which is
operated by the Association of Universities for Research in Astronomy, under
NASA contract NAS5-26555. Work performed at LLNL is supported by the DOE under
contract W7405-ENG-48.  Work performed by the Center for Particle Astrophysics
personnel is supported in part by the Office of Science and Technology Centers
of NSF under cooperative agreement AST-8809616. DM is also supported by
Fondecyt 1990440.  CWS thanks the Packard Foundation for the generous support.
WJS is supported by a PPARC Advanced Fellowship.  CAN is supported in part by
a NPSC Graduate Fellowship.  TV and KG were supported in part by the DOE under
grand DEF0390-ER 40546.  TV was supported in part by an IGPP grant.

\appendix


\clearpage



\clearpage





\clearpage

\begin{deluxetable}{lllccl}
\label{table1}
\tablewidth{0pt}
\tablecaption{Summary of Observations}
\tablehead{
\colhead {Event} & \colhead{RA} & \colhead {DEC} & \colhead {V Exposure Times}
&\colhead{I Exposure Times} &\colhead{Obs Date}}
\startdata
LMC-1   & 05:14:44.50 & -68:48:00.00 & 4X400s & 40X500s & 1997-12-16 \\
LMC-4   & 05:17:14.60 & -70:46:59.00 & 4X400s &  2X500s & 1998-08-19 \\
LMC-5   & 05:16:41.10 & -70:29:18.00 & 4X400s &  2X500s & 1999-05-13 \\
LMC-6   & 05:26:14.00 & -70:21:15.00 & 4X400s &  2X500s & 1999-08-26 \\
LMC-7   & 05:04:03.40 & -69:33:19.00 & 4X400s &  2X500s & 1999-04-12 \\
LMC-8   & 05:25:09.40 & -69:47:54.00 & 4X400s &  2X500s & 1999-03-12 \\
LMC-9   & 05:20:20.30 & -69:15:12.00 & 4X400s &  2X500s & 1999-04-13 \\
LMC-14  & 05:34:44.40 & -70:25:07.00 & 4X500s &  4X500s & 1997-05-13 \\
\enddata
\end{deluxetable}

\begin{deluxetable}{lrr}
\label{table1}
\tablewidth{0pt}
\tablecaption{Photometry of Microlensing Source Stars}
\tablehead{
\colhead {Event} & \colhead{$V$} & \colhead {$V-I$} }
\startdata
LMC-1 & $19.782\pm0.003$ & $1.167\pm0.004$ \\
LMC-4 & $21.331\pm0.008$ & $0.502\pm0.009$ \\
LMC-5 & $21.016\pm0.096$ & $0.677\pm0.122$ \\
LMC-6 & $20.041\pm0.004$ & $0.471\pm0.007$ \\
LMC-7 & $21.993\pm0.013$ & $0.672\pm0.022$ \\
LMC-8 & $20.195\pm0.004$ & $0.203\pm0.009$ \\
LMC-9a& $21.137\pm0.007$ & $0.991\pm0.011$ \\
LMC-9b& $21.250\pm0.008$ & $1.002\pm0.012$ \\
LMC-14& $19.467\pm0.002$ & $0.106\pm0.004$ \\
\enddata
\end{deluxetable}

\end{document}